\documentclass[conference]{IEEEtran}
\IEEEoverridecommandlockouts

\usepackage{cite}
\usepackage{amsmath,amssymb,amsfonts}
\usepackage{algorithmic}
\usepackage{algorithm}
\usepackage{array}
\usepackage{textcomp}
\usepackage{stfloats}
\usepackage{url}
\usepackage{verbatim}
\usepackage{subcaption}
\usepackage{graphicx}
\usepackage{bm}
\usepackage[dvipsnames]{xcolor}
\usepackage{balance}
\usepackage{tikz}
\usepackage{pgfplots}
\pgfplotsset{compat=newest}
\usetikzlibrary{fit, backgrounds, positioning, shapes, arrows, calc, arrows.meta}
\usepackage{fancyhdr}
\newcommand{\preprintheader}{{\footnotesize\scshape Preprint, \today}}
\fancypagestyle{plain}{%
  \fancyhf{}%
  \fancyhead[R]{\preprintheader}%
}

\begin{document}

\title{Characterizing Nonlinearities in IM-DD Links via the Best Linear Approximation: Distortion Analysis and Modulation Optimization}

\author{Sebastian Fraga Fern\'{a}ndez\textsuperscript{\textdagger,*}, Leonardo Minelli\textsuperscript{\textdaggerdbl}, Fernando de Bernardinis\textsuperscript{\textdaggerdbl},\\
Felipe Villenas\textsuperscript{\textdagger}, Yunus Can G\"{u}ltekin\textsuperscript{\textdagger}, Alex Alvarado\textsuperscript{\textdagger} \\
\textit{\textsuperscript{\textdagger}Department of Electrical Engineering, Eindhoven University of Technology} \\
\textit{\textsuperscript{\textdaggerdbl}Marvell Technology, Pavia, Italy} \\
\textsuperscript{*}s.fraga.fernandez@tue.nl
}

\maketitle
\thispagestyle{fancy}  % force the fancy header on the title page (IEEEtran conference class defaults to IEEEtitlepagestyle here)

\begin{abstract}
We use the best linear approximation (BLA) to characterize nonlinearities in IM-DD links. The orthogonal nonlinear distortion is shown to be non-Gaussian. The BLA is used to approximate the optimum modulation depth for different equalizers.
\end{abstract}

\begin{IEEEkeywords}
Mach-Zehnder modulator, best linear approximation, orthogonal component, pulse amplitude modulation, nonlinear distortion, intensity modulation and direct detection
\end{IEEEkeywords}

\section{Introduction}
Intensity modulation and direct detection (IM-DD) remain the most widely adopted technology for short-reach optical interconnects due to their simple implementation, low cost, and favorable power efficiency \cite{Zhou_JLT}. The rapid growth of data-centric applications is driving these links towards operation at higher baud rates and more advanced modulation formats \cite{Che_JLT}, thereby accentuating transceiver nonidealities. Nonlinear distortion from electro-optical (EO) modulators is one of the major impairments in IM-DD systems.

Radio frequency metrics used to quantify nonlinear distortion are based on tone excitations. While suitable for analog photonic links \cite{Luo_JLT, Luo_IEEE_MTTs, Anderson_IEEE_Photonics}, these metrics provide limited insight for IM-DD links. This is because the waveforms used in these systems differ from tone excitations, and nonlinear distortion depends not only on the system itself, but also on the statistical properties of the excitation \cite{Gharaibeh}.

In optical communications, the best linear approximation (BLA) has been used in \cite{Su_OECC} as a tool to characterize nonlinear distortions. The BLA\footnote{The BLA has been used under the names of orthogonal component analysis \cite{Li_JLT} and orthogonal decomposition \cite{Lin_JLT, Li_JLT_2}, and is regarded as a benchmark for characterizing nonlinear distortion \cite{Yang_OFC, Tao_OFC, Tao_JLT, Ye_communications_engineering, Li_JLT}.} of a system is defined as the linear time-invariant (LTI) system whose output minimizes the mean-squared error (MSE) relative to the output of the original system [\citen{BLA}, Ch.~5]. In \cite{Su_OECC}, the BLA was used to derive an equivalent representation of a link, from which system performance, in terms of bit error rate (BER) or Q-factor, can be estimated accurately.

In this paper, we use the BLA to characterize nonlinear distortion in a Mach--Zehnder modulator (MZM)-based IM-DD link. The main contributions of this work are two. Firstly, we study the statistical properties of the orthogonal nonlinear distortion and show that it is not Gaussian, as sometimes assumed in the literature [\citen{BLA}, Ch.~5]. Secondly, we show that the BLA can be used to obtain an approximation of the optimal MZM drive voltage swing for different equalizer structures.

\section{System Model}
\begin{figure*}[t]
    \centering
    \input{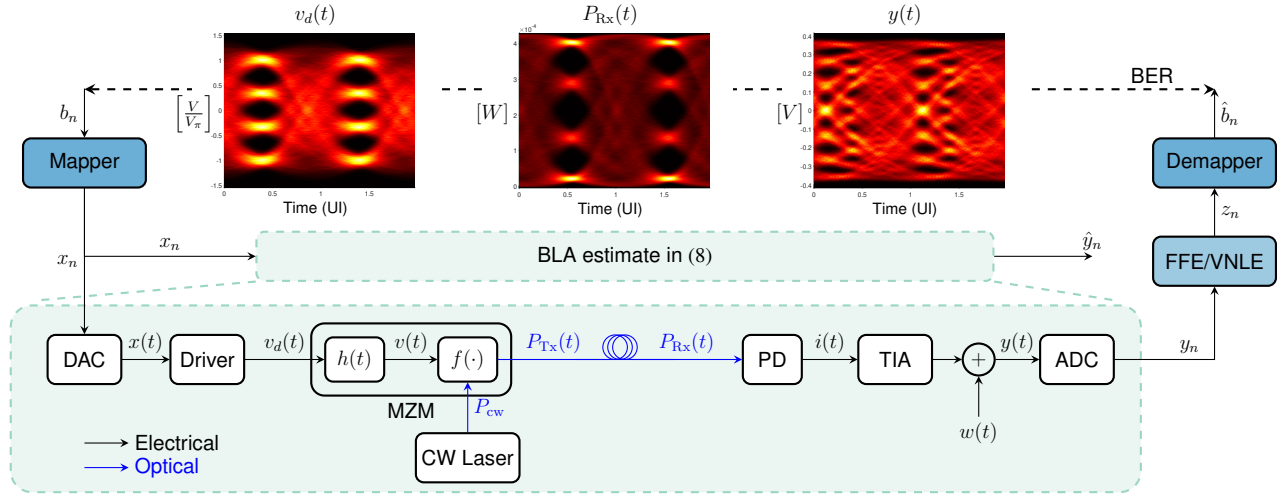}
    \caption{IM-DD link under consideration. The BLA models the transceiver chain as an LTI system with AWGN. The displayed eye diagrams were obtained with $k = 0.35$ and $I = 5~\mu A$.}
    \label{System_model}
\end{figure*}

The considered IM-DD system is shown in Fig.~\ref{System_model}. Randomly generated bits $b_n$ are Gray-mapped to 4-ary pulse amplitude modulation (PAM-4) symbols $x_n \in \{ \pm 1/3, \pm 1 \}$. The digital-to-analog converter (DAC) is modeled as a zero-order hold, producing the continuous-time waveform
\begin{equation}
x(t) = \sum_{n} x_n p(t - nT_s),
\end{equation}
where $T_s = 1/R_s$ is the symbol duration determined by the symbol rate $R_s$, and $p(t)$ is a rectangular pulse of unit amplitude and duration $T_s$.

The voltage $x(t)$ is fed into an LTI model of a commercial driver, producing an output $v_d(t)$. We model the MZM as a cascade of two systems \cite{Ghione_2009, Shi_2018}. First, a first-order low-pass filter with impulse response $h(t)$, unity passband gain, and bandwidth $BW_{1}$ transforms the applied drive voltage $v_d(t)$ into the effective modulation voltage $v(t)$. Second, the static EO response of an ideal MZM biased at $-V_\pi/2$ is applied. Assuming an effective index change proportional to $v(t)$, the transmitted optical power is
\begin{equation}
P_{\mathrm{Tx}}(t) = f(v(t)), \,\,
f(x) \triangleq \frac{P_{\mathrm{cw}}}{2} \left[ 1 + \sin \left( \pi \frac{x}{V_\pi} \right) \right].
\label{eq:MZM_transmission}
\end{equation}
where $P_{\mathrm{cw}}$ is the power, from a continuous-wave (CW) source, that is coupled into the MZM, and where $V_\pi$ is the half-wave voltage of the MZM.

The effective voltage $v(t)$ can be expressed as
\begin{equation}
v(t) = \sum_n x_n u(t-nT_s),
\end{equation}
where $u(t)$ is the response of the cascade of the driver and the system with impulse response $h(t)$, to an input $p(t)$.

We define the modulation depth $k$ as
\begin{equation} \label{eq:modulation_depth}
k \triangleq \frac{\textrm{max}|u(t)|}{V_{\pi}},
\end{equation}
which quantifies the level of excursion in the MZM electro-optic transmission given by \eqref{eq:MZM_transmission}. A high $k$ results in a high optical modulation amplitude (OMA), but it also leads to stronger nonlinear distortion. Without loss of generality, in this paper we use $V_\pi  =1$ V.

$P_{\mathrm{Tx}}(t)$ is launched into the fiber, which is modeled as a linear, nondispersive channel with attenuation. The received optical power $P_{\mathrm{Rx}}(t)$ is converted to a photocurrent $i(t)$ by a photodiode (PD), modeled as a first-order low-pass filter with bandwidth $BW_{2}$ and responsivity $\mathcal{R}$. The current $i(t)$ is AC-coupled into a transimpedance amplifier (TIA), modeled as a fourth-order Bessel filter with bandwidth $BW_{3}$ and gain $Z$.

The output-referred TIA noise $w(t)$ is modeled as additive white Gaussian noise (AWGN) with variance $\sigma_w^2 = (Z I)^2$ and
single-sided power spectral density (PSD) $N_0 = {2\sigma_w^2}/{(R_s N_s)}$, where $N_s$ is the simulation oversampling factor
and $I$ is the root mean square (RMS) input-referred TIA noise current. The TIA output $y(t)$ is sampled at the symbol rate by an analog-to-digital converter (ADC) and processed by either an 8-tap feed-forward equalizer (FFE) or a third-order Volterra nonlinear equalizer (VNLE) \cite{VNLE} with 8 and 35 taps in the first- and third-order kernels, respectively. Both equalizers are adapted using the least mean squares (LMS) algorithm. For each equalizer, the ADC sampling phase was swept to achieve the minimum BER. The system parameters are listed in Table \ref{tab:sim_params}.

\begin{table}[t!]
\centering
\caption{System Parameters}
\label{tab:sim_params}
\footnotesize
\begin{tabular}{lc}
\hline
Parameter & Value \\
\hline
Symbol rate $R_s$ & 112~GBd\\
Input power $P_{\mathrm{cw}}$ & 0.5~mW\\
EO bandwidth $BW_{1}$ & 100.8 GHz \\
Fiber length & 2~km\\
Attenuation coefficient & 0.35 dB/km\\
PD bandwidth $BW_{2}$ & 56 GHz \cite{Photodiode}\\
PD responsivity $\mathcal{R}$ & 0.9~A/W  \cite{Photodiode}\\
TIA bandwidth $BW_{3}$ & 56 GHz\\
TIA gain $Z$ & 66~dB$\Omega$\\
%Noise PSD & $\frac{2(Z I)^2}{F_s}$ $V^2$/Hz\\
Oversampling $N_s$ & $32$ Sa/sym \\
\noalign{\smallskip}
\hline
\end{tabular}
\end{table}

\section{BLA of the IM-DD Link}

The BLA of order $M$ of a system, for an input $x_n$, is defined as the impulse response of an $M$-tap finite impulse response (FIR) filter that minimizes the mean-square error
\begin{equation}\label{eq.mse}
J = \mathbb{E}\big\{ (y_n - \hat y_n)^2 \big\},
\end{equation}
where $\hat y_n$ is the filter output. The residue
\begin{equation}
e_n = y_n - \hat y_n
\end{equation}
satisfies the orthogonality conditions
\begin{equation}
\mathbb{E}\{ e_n \, x_{n-m} \} = 0, \quad m = 0,1,\dots,M-1.
\label{eq:orthogonality}
\end{equation}
The linear BLA estimate can be written as
\begin{equation}\label{eq:bla}
\hat y_n = K\, x_{n-q} + g_n,
\end{equation}
where $K$ is the main tap of the BLA, $q$ is the delay introduced by the link, and $g_n$ is linear inter-symbol interference. Fig.~\ref{System_model} shows the application of the BLA, where the complete IM-DD link is replaced by a discrete-time model given by \eqref{eq:bla}.

The received samples can be expressed as
\begin{equation}\label{eq:yn}
y_n = \hat y_n + w_n + d_n,
\end{equation}
where $w_n$ is the noise component from $w(t)$. In \eqref{eq:yn}, the term $d_n$ is the component of the noiseless output $y_n-w_n$ orthogonal to the subspace spanned by $x_n, \ldots, x_{n-M+1}$. In other words, $d_n$ is the component of $y_n$, in the absence of noise, that cannot be captured by the used FIR filter model. Since $d_n$ and $w_n$ are zero-mean and uncorrelated, the MSE in \eqref{eq.mse} decomposes as
\begin{equation}\label{eq.J2}
J = \mathbb{E}\{d_n^2\} + \mathbb{E}\{w_n^2\},
\end{equation}
which shows that $d_n$ can be treated as an additive noise term.

The term $d_n$ consists of two contributions: \emph{i}) orthogonal nonlinear distortion (ONL) and \emph{ii}) a projection error arising from the finite dimensionality of the regression subspace. For sufficiently large $M$, the latter becomes negligible, and $d_n$ effectively consists exclusively of ONL. Throughout this work, the BLA was computed with $M$ large enough to ensure the latter contribution is negligible. In the next section, we provide a detailed statistical characterization of $d_n$ .

\section{Numerical Results}

\begin{figure*}[t!]
\centering
\begin{tikzpicture}
\begin{axis}[
    grid=major,
    major grid style={solid, gray!20},
    minor tick num=10,
    tick align=inside,
    major tick length=6pt,
    minor tick length=2pt,
    tick style={black, thin},
    tick label style={font=\footnotesize},
    axis background/.style={fill=white},
    legend style={
    at={(0.99,1)},
    anchor=north east,
    legend columns=4,
    column sep=0.5em,
    font=\footnotesize,
    inner sep=1pt},
    width=1.0\textwidth,
    height=0.27\textwidth,
    xmin=-1.2, xmax=1.2,
    ymin=-0.1, ymax = 18
]
\addplot [black, solid, thick] table {data/pdf_x_minus1.txt} node[pos=0.65, xshift=5pt, right, font=\footnotesize, align=left] {$\mu = -0.985$\\$\sigma^2 = 0.004$};
\addlegendentry{$p\left( \frac{d_n}{K} | x_{n-q} = -1\right)$}

\addplot [blue, solid, thick] table {data/pdf_x_minus1_3.txt}
node[pos=0.64, right, font=\footnotesize, align=left] {$\mu = -0.375$\\$\sigma^2 = 0.003$};
\addlegendentry{$p\left( \frac{d_n}{K}| x_{n-q} = -\frac{1}{3}\right)$}

\addplot [PineGreen, solid, thick] table {data/pdf_x_plus1_3.txt}
node[pos=0.64, left, xshift=-9pt, font=\footnotesize, align=left] {$\mu = 0.375$\\$\sigma^2 = 0.003$};
\addlegendentry{$p\left( \frac{d_n}{K} | x_{n-q} = \frac{1}{3}\right)$}

\addplot [red, solid, thick] table {data/pdf_x_plus1.txt}
node[pos=0.65, left, xshift=-10pt, font=\footnotesize, align=left] {$\mu = 0.985$\\$\sigma^2 = 0.004$};
\addlegendentry{$p\left(\frac{d_n}{K} | x_{n-q} = 1\right)$}

\end{axis}
\end{tikzpicture}
\caption{PDF of ${d_n}/{K}$ conditioned on the transmitted symbol $x_{n-q}$, for $k = 0.35$.}
\label{fig:PDFs}
\end{figure*}
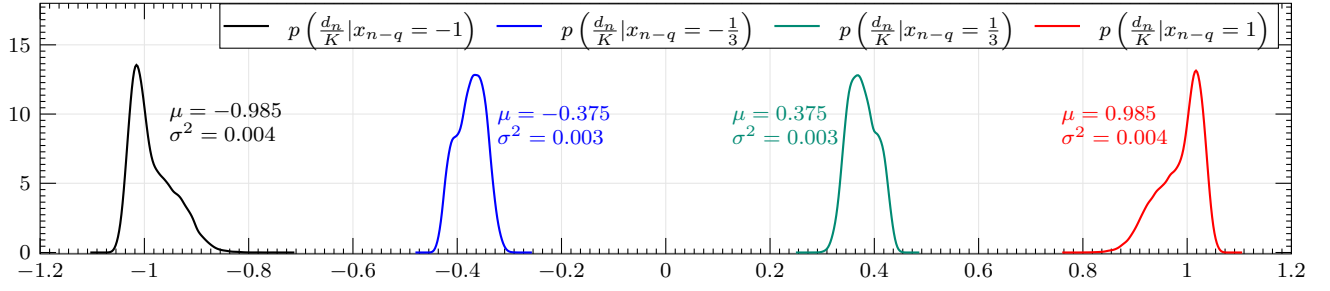

Fig. \ref{fig:PDFs} shows the probability density function (PDF) of $d_n$, normalized by $K$, conditioned on the transmitted symbols $x_{n-q}\in\{\pm 1/3,\pm 1\}$. This figure shows that the conditional PDFs are not Gaussian. Moreover, the PDFs conditioned on opposite symbol levels are symmetric with respect to zero. This follows from the odd nature of the second term in the modulator transmission $f(x)$ in \eqref{eq:MZM_transmission}, and the linearity of the remaining blocks. Although $d_n$ is zero mean, the means ($\mu$) of the conditional PDFs are not equal to the respective transmitted symbol. This is because the BLA is a linear approximation for which the optimization is carried out over the entire input distribution, rather than a fit for each PAM-4 level. The compression experienced at each PAM-4 level creates a deterministic residual (a local DC component) at that specific symbol's location. This results in a shift of the conditional PDF mean for that symbol. Because $f(x)$ is odd and the constellation is symmetric, the negative mean shift at $+1$ is perfectly balanced by the positive mean shift at $-1$. The same occurs for the inner symbols $\pm 1/3$. Thus, these individual DC components appear in the conditional PDFs even though they cancel out in the total mean.

% Probability density functions

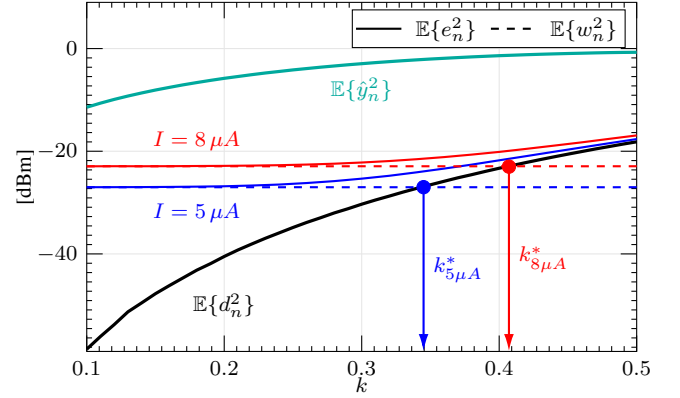
\begin{figure}[t]
\centering
\begin{tikzpicture}
\begin{axis}[
    xlabel={\footnotesize $k$},
    xlabel style={yshift=2mm},
    ymin=-59, ymax=9,
    ylabel={\footnotesize [dBm]},
    ylabel style={yshift=-2mm},
    tick label style={font=\footnotesize},
    xmin=0.1, xmax=0.5,
    xtick={0.1, 0.2, 0.3, 0.4, 0.5},
    grid=major,
    major grid style={solid, gray!20},
    minor tick num=10,
    tick align=inside,
    major tick length=6pt,
    minor tick length=2pt,
    tick style={black, thin},
    axis background/.style={fill=white},
    legend style={at={(0.98,0.98)}, anchor=north east, legend columns=2, font=\footnotesize, inner sep=1pt, column sep=3pt},
    width=\columnwidth,
    height=0.7\columnwidth,
    clip mode=individual
]

% P_yc: Solid line, filled circle markers
\addplot [Emerald, solid, very thick, forget plot] table {data/power_yhat.txt};
\node[Emerald, font=\footnotesize] (label_yhat) at (axis cs:0.30,-8) {$\mathbb{E}\{\hat{y}_n^2\}$};

% P_ynl: Solid line
\addplot [black, solid, very thick, forget plot] table {data/power_dn.txt};
\node[font=\footnotesize] (label_dn) at (axis cs:0.20,-50) {$\mathbb{E}\{d_n^2\}$};

% w_n: dashed, 5uA
\addplot [blue, dashed, thick, forget plot] table {data/power_wn_5uA.txt};

% e_n: solid, 5uA
\addplot [blue, solid, thick, forget plot] table {data/power_wn_8uA.txt};
\node[blue, font=\footnotesize] at (axis cs:0.18,-32) {$I= 5\,\mu A$};

% w_n: dashed, 8uA
\addplot [red, dashed, thick, forget plot] table {data/power_en_5uA.txt};

% e_n: solid, 8uA
\addplot [red, solid, thick, forget plot] table {data/power_en_8uA.txt};
\node[red, font=\footnotesize] at (axis cs:0.18,-18) {$I= 8\,\mu A$};

% Legend entries for line styles only
\addlegendimage{solid, thick, black}
\addlegendentry{$\mathbb{E}\{e_n^2\}$}
\addlegendimage{dashed, thick, black}
\addlegendentry{$\mathbb{E}\{w_n^2\}$}

% --- Blue Arrow ---
\addplot[blue, only marks, mark=*, mark size=2.5pt, forget plot] coordinates {(0.345,-27)};
\draw[-{Latex[length=2.3mm, width=1.3mm]}, blue, thick]
    (axis cs:0.345,-27) -- (axis cs:0.345,-59)
    node[midway, right, font=\footnotesize] {$k^*_{5\mu A}$};

% --- Red Arrow ---
\addplot[red, only marks, mark=*, mark size=2.5pt, forget plot] coordinates {(0.4071,-23)};
\draw[-{Latex[length=2.3mm, width=1.3mm]}, red, thick]
    (axis cs:0.4071,-23) -- (axis cs:0.4071,-59)
    node[midway, right, font=\footnotesize] {$k^*_{8\mu A}$};

\end{axis}
\end{tikzpicture}
\caption{Evolution of different powers vs. modulation depth $k$. The critical values of $k$ are shown with filled circles.}
\label{fig:BLA_power_vs_k}
\end{figure}

Fig.~\ref{fig:BLA_power_vs_k} shows the evolution of the power of $\hat{y}_n$, $d_n$, $e_n$ and $w_n$ as a function of $k$. The latter two are plotted for two different TIA noise currents: $I= 5~\mu A$ and $I = 8~\mu A$.
In both cases, for small values of $k$, $\mathbb{E}\{w_n^2\}$ dominates over $\mathbb{E}\{d_n^2\}$, making the system AWGN-limited. Increasing $k$ results in a quasi-linear increase of $P_{\mathrm{Tx}}(t)$ swing (see \eqref{eq:MZM_transmission}). At a critical $k = k^*_I$, $\mathbb{E}\{d_n^2\}$ matches $\mathbb{E}\{w_n^2\}$. For larger values of \(k\) the system becomes limited by nonlinear distortion. $\mathbb{E}\{y_n^2\}$ continues to increase with \(k\), but the rate of increase becomes noticeably slower, reflecting the onset of gain compression.

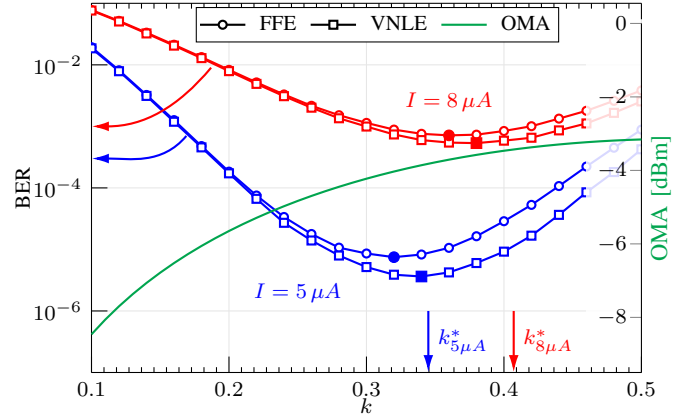
\begin{figure}[t]
\centering
\begin{tikzpicture}
\begin{axis}[
    axis y line*=left,
    xlabel={\footnotesize $k$},
    ylabel={\footnotesize BER},
    ylabel style={yshift=-2mm},
    xlabel style={yshift=2mm},
    tick label style={font=\footnotesize},
    xmin=0.1, xmax=0.5,
    ymax = 0.1,
    ymin=1e-7,
    ymode=log,
    xtick={0.1, 0.2, 0.3, 0.4, 0.5},
    grid=major,
    major grid style={solid, gray!20},
    minor tick num=10,
    tick align=inside,
    major tick length=6pt,
    minor tick length=2pt,
    tick style={black, thin},
    axis background/.style={fill=white},
    legend style={at={(0.52,0.99)}, anchor=north, legend columns=3, font=\footnotesize, inner sep=1pt, column sep=3pt},
    width=\columnwidth,
    height=0.73\columnwidth,
    clip mode=individual,
    set layers=standard,
]

%% -------------------- 5uA -------------------- %%
% FFE: Solid line, filled circle markers
\addplot [blue, solid, mark=*, mark size=1.5pt, mark options={fill=white}, thick, forget plot,mark layer=like plot] table {data/ber_ffe_5uA.txt};

% VNLE: Solid line, filled square markers
\addplot [blue, solid, mark=square*, mark size=1.5pt, mark options={fill=white}, thick, forget plot,mark layer=like plot] table {data/ber_vnle_5uA.txt};

\node[blue, font=\footnotesize] at (axis cs:0.25,2e-6) {$I= 5\,\mu A$};
%% --------------------------------------------- %%

%% -------------------- 8uA -------------------- %%
% FFE: Solid line, filled circle markers
\addplot [red, solid, mark=*, mark size=1.5pt, mark options={fill=white}, thick, forget plot,mark layer=like plot] table {data/ber_ffe_8uA.txt};

% VNLE: Solid line, filled square markers
\addplot [red, solid, mark=square*, mark size=1.5pt, mark options={fill=white}, thick, forget plot,mark layer=like plot] table {data/ber_vnle_8uA.txt};

\node[red, font=\footnotesize] at (axis cs:0.36, 2.5e-3) {$I= 8\,\mu A$};
%% --------------------------------------------- %%

% Filled markers at BER minima
\addplot [blue, only marks, mark=*, mark size=2pt, mark options={fill=blue}, forget plot] coordinates {(0.32, 7.487e-6)};
\addplot [blue, only marks, mark=square*, mark size=2pt, mark options={fill=blue}, forget plot] coordinates {(0.34, 3.632e-6)};
\addplot [red, only marks, mark=*, mark size=2pt, mark options={fill=red}, forget plot] coordinates {(0.36, 7.141e-4)};
\addplot [red, only marks, mark=square*, mark size=2pt, mark options={fill=red}, forget plot] coordinates {(0.38, 5.342e-4)};

% Arrows relating BER curves with left axis
\draw[-{Latex[length=2.3mm, width=1.3mm]}, blue, thick] (axis cs:0.169, 7e-4) to[out=230, in=0] (axis cs:0.1, 3e-4);
\draw[-{Latex[length=2.3mm, width=1.3mm]}, red, thick] (axis cs:0.187, 9e-3) to[out=230, in=0] (axis cs:0.1, 1e-3);

% Legend: line styles only
\addlegendimage{solid, mark=*, mark size=1.5pt, mark options={fill=white}, thick, black}
\addlegendentry{FFE}
\addlegendimage{solid, mark=square*, mark size=1.5pt, mark options={fill=white}, thick, black}
\addlegendentry{VNLE}
\addlegendimage{solid, thick, Green}
\addlegendentry{OMA}

% Arrows pointing to optimum k values
% --- Blue Arrow ---
% \addplot[blue, only marks, mark=*, mark size=2pt, forget plot] coordinates {(0.345,1e-6)};
\draw[-{Latex[length=2.3mm, width=1.3mm]}, blue, thick]
    (axis cs:0.345,1e-6) -- (axis cs:0.345,1e-7)
    node[midway, right, font=\footnotesize] {$k^*_{5\mu A}$};

% --- Red Arrow ---
% \addplot[red, only marks, mark=*, mark size=2pt, forget plot] coordinates {(0.4071,1e-6)};
\draw[-{Latex[length=2.3mm, width=1.3mm]}, red, thick]
    (axis cs:0.4071,1e-6) -- (axis cs:0.4071,1e-7)
    node[midway, right, font=\footnotesize] {$k^*_{8\mu A}$};

% Transparent rectangle for second y-axis
%\begin{pgfonlayer}{axis ticks}
\draw[fill=white,draw=none,opacity=0.85] (axis cs:0.46, 5e-7) rectangle (0.5,3e-2);
%\end{pgfonlayer}

\end{axis}

\begin{axis}[
    width=\columnwidth,
    height=0.7\columnwidth,
    xmin=0.1, xmax=0.5,
    axis y line*=right,
    axis x line=none,
    axis y line node/.style={Green},
    %ytick style={Green},
    ytick align=inside,
    yticklabel style={anchor=east,xshift=-2pt},
    ylabel style={Green},
    ylabel={\small OMA [dBm]},
    ymin=-9.5, ymax=0,
    ytick={-8,-6,...,0},
    font=\footnotesize
]
    \addplot [Green, thick] table {data/OMA.txt};
\end{axis}

\end{tikzpicture}
\caption{BER and OMA vs. modulation depth $k$. Optimum BER values are shown with filled markers.}
\label{fig:BLA_BER_vs_k}
\end{figure}

Fig.~\ref{fig:BLA_BER_vs_k} displays the OMA, and the BER for the two equalizers considered in this work, as functions of $k$ . For small $k$, the system is AWGN-limited, and hence, the FFE and VNLE perform nearly identically. As $k$ increases, the BER starts decreasing due to the increased signal swing, while the contribution of $d_n$ remains secondary compared to $w_n$. For both values of $I$, both equalizers achieve a BER minimum at approximately the respective value of $k^*$ (from Fig.~\ref{fig:BLA_power_vs_k}). Beyond this point, the BER rises again due to $d_n$ becoming dominant, with the VNLE outperforming the FFE since the system is now limited by nonlinear distortion.

% Power decomposition

% BER vs k plot

\section{Conclusions}
We used the BLA to characterize the nonlinear distortion in an MZM-based IM-DD link. For a given transmit power and two different equalizers, we showed that the minimum BER is achieved at approximately the modulation depth that perfectly balances the AWGN and the ONL powers. Future work includes an experimental validation of the technique presented in this paper.

\section*{Acknowledgements}
This research is part of the project COmplexity-COnstrained LIght-coherent optical links \mbox{(COCOLI)} funded by Holland High Tech $|$ TKI HSTM via the PPS allowance scheme for public-private partnerships.

\bibliographystyle{IEEEtran}
\bibliography{references}

\end{document}